\documentclass[aps,prl,preprint,groupedaddress]{revtex4-1}
 \usepackage{graphicx}
 \usepackage{dcolumn}
 \usepackage{bm}
 \usepackage{amsmath}
 \usepackage{color}
 \usepackage{amssymb}
 \usepackage{mathrsfs}
 \usepackage{MnSymbol}
 \usepackage{wasysym}
 \usepackage{gensymb}
 \usepackage{framed,color}
 \usepackage{enumitem}

\begin{document}

\title{Superconductivity and local stacking polymorphism in LaNiGa$_2$ thin films}

\author{Pengcheng Wang$^1$, Qingxiao Wang$^2$, Yameng Wu$^1$, Zixuan Huang$^1$, C. T. Lam$^3$, Zheyu Wang$^1$, Swee K. Goh$^1$, Xiaoying Xu$^3$, Yufan Li$^1$}

\affiliation{$^1$Department of Physics, The Chinese University of Hong Kong, Shatin, Hong Kong SAR, China}
\affiliation{$^2$Department of Electrical and Computer Engineering, King Abdullah University of Science and Technology, Thuwal 23955-6900, Saudi Arabia}
\affiliation{$^3$ Quantum Science Center of Guangdong-Hong Kong-Macao Greater Bay Area, Shenzhen, Guangdong, China}

\date{\today}

\begin{abstract}

LaNiGa$_2$ has attracted considerable interest as a candidate unconventional superconductor, particularly due to its reported time-reversal symmetry breaking and nonsymmorphic crystal structure hosting symmetry-protected electronic band degeneracies.
Here, we report the growth and characterization of superconducting LaNiGa$_2$ thin films on MgO substrates by magnetron sputtering.
X-ray diffraction reveals highly crystalline, (010)-oriented films with well-defined in-plane alignment and rotational twinning.
Electrical transport measurements establish superconductivity with a transition temperature of approximately 1 K.
Atomic-resolution transmission electron microscopy reveals local variations in crystallographic orientation and interlayer stacking, with a predominantly \textit{Cmcm} structure containing minority \textit{Cmmm} regions.
The coexistence of these closely related structures reveals local stacking polymorphism that is difficult to distinguish using conventional X-ray diffraction.
These findings establish a superconducting LaNiGa$_2$ thin-film platform and provide new insight into the local crystal structure of a material whose electronic properties and superconducting pairing remain under active investigation.

\end{abstract}

\maketitle

\begin{text}
LaNiGa$_2$ has emerged as an intriguing candidate for unconventional superconductivity in a material with nontrivial topological electronic bands, combining evidence for time-reversal symmetry breaking (TRSB) with symmetry-protected Dirac states at the Fermi level.
Zero-field muon-spin relaxation ($\mu$SR) measurements revealed spontaneous internal magnetic fields below the superconducting transition temperature, providing evidence for TRSB \cite{hillier_nonunitary_2012}.
Subsequent measurements of heat capacity, London penetration depth, and upper critical field suggested fully gapped two-gap superconductivity \cite{weng_two-gap_2016,sundar_gap_2024}.
These observations motivated the proposal of an internally antisymmetric nonunitary triplet (INT) pairing state, in which equal-spin Cooper pairs form between electrons occupying different orbitals, allowing even-parity spin-triplet superconductivity \cite{weng_two-gap_2016,ghosh_quantitative_2020}.
The possibility of such an unconventional pairing state, together with its unusual electronic structure, makes LaNiGa$_2$ a promising platform for exploring the interplay between superconductivity and band topology.

The topological electronic structure of LaNiGa$_2$ is closely tied to its crystalline symmetry.
Originally assigned to the symmorphic \textit{Cmmm} space group, its crystal structure was subsequently reassigned to the nonsymmorphic \textit{Cmcm} space group through single-crystal X-ray diffraction \cite{badger_dirac_2022}.
This reassignment has profound implications for its electronic structure, as the nonsymmorphic symmetry enforces band degeneracies that give rise to Dirac lines and a Dirac loop at the Fermi level \cite{badger_dirac_2022,quan_nonsymmorphic_2022}.
These features have received further experimental support from angle-resolved photoemission spectroscopy and de Haas-van Alphen measurements \cite{staab_symmetry-enforced_2024,li_fermiology_2025}.
The symmetry-enforced band degeneracies also provide favorable conditions for interband superconducting pairing, suggesting a possible connection between the topological electronic structure and unconventional superconductivity.
Remarkably, the \textit{Cmcm} and \textit{Cmmm} structures differ primarily by the relative displacement of neighboring atomic layers, yet their distinct crystalline symmetries have substantially different consequences for the electronic band structure.
The local arrangement of these layers is therefore particularly relevant for understanding the electronic properties of LaNiGa$_2$.

Experimental studies of superconducting LaNiGa$_2$ have predominantly focused on bulk polycrystalline samples and single crystals.
The realization of superconducting thin films would provide opportunities to investigate this material in reduced dimensions and enable the fabrication of superconducting devices and heterostructures.
Here, we report the growth of highly crystalline LaNiGa$_2$ thin films on MgO substrates by dual-source dc magnetron sputtering.
The films exhibit a pronounced (010) crystallographic orientation with in-plane rotational twinning and superconductivity near 1 K.
Atomic-resolution transmission electron microscopy reveals distinct crystallographic orientations and local variations in interlayer stacking, with the dominant nonsymmorphic \textit{Cmcm} structure accompanied by minority \textit{Cmmm}-like regions.
These findings establish a superconducting LaNiGa$_2$ thin-film platform and provide new insights into its local crystalline structure, with implications for the symmetry-protected electronic states proposed for this material.

LaNiGa$_2$ thin films were grown on MgO(001) substrates by co-sputtering La and a Ni–Ga alloy target (Ni:Ga = 1:2, atomic ratio) in an ultrahigh-vacuum magnetron sputtering system with a base pressure of $3\times10^{-9}$ Torr.
The substrate temperature was maintained at 700~$^\circ$C during deposition under an Ar working pressure of 3 mTorr.
The sputtering powers were adjusted to obtain the desired LaNiGa$_2$ stoichiometry. 
The crystallographic structure and orientation of the films were examined by X-ray diffraction (XRD).
As shown in Fig.~1(b), the $\theta$--$2\theta$ scan exhibits a series of pronounced LaNiGa$_2$ (0$k$0) reflections, together with the MgO (002) substrate reflection.
No diffraction peaks from secondary crystalline phases are detected.
The observed reflection sequence is consistent with the reported bulk LaNiGa$_2$ structure \cite{badger_dirac_2022,li_fermiology_2025,li_pressure-induced_2026}.
The exclusive observation of (0$k$0) reflections indicates a strong preferred out-of-plane orientation, with the crystallographic $b$ axis perpendicular to the substrate surface.
The in-plane crystallographic orientation was further examined by azimuthal $\phi$ scans of the LaNiGa$_2$ (150) reflection.
As shown in Fig.~1(c), four sharp peaks separated by 90$^\circ$ are observed, with an azimuthal full width at half maximum (FWHM) of approximately 1.46$^\circ$, comparable to that of the MgO substrate.
The film reflections exhibit a well-defined angular relationship with the MgO substrate reflections, indicating a high degree of in-plane crystallographic alignment and a small mosaic spread.

The two structural models, \textit{Cmcm} and \textit{Cmmm}, are compared in Fig.~1(a).
Despite their different symmetries, they have very similar lattice parameters.
The reported lattice parameters are $a=4.278$~\AA, $b=17.436$~\AA, and $c=4.271$~\AA~for the centrosymmetric orthorhombic \textit{Cmmm}, and $a=4.273$~\AA, $b=17.412$~\AA, and $c=4.268$~\AA~for the nonsymmorphic \textit{Cmcm} \cite{zeng_superconductivity_2002,badger_dirac_2022,li_pressure-induced_2026}.
In both structures, the $a$ and $c$ lattice parameters are nearly identical.
With the crystallographic $b$ axis normal to the substrate, the $a$ and $c$ axes lie within the film plane.
A single orthorhombic domain would produce twofold azimuthal symmetry, whereas the observed fourfold symmetry indicates the presence of two in-plane crystallographic variants related by a 90$^\circ$ rotation.
Such rotational twinning can be understood from the near equivalence of the LaNiGa$_2$ $a$ and $c$ axes and the fourfold symmetry of the MgO(001) substrate.
The two variants correspond to the orientation relationships $[100]_{\mathrm{LaNiGa_2}}\parallel[100]_{\mathrm{MgO}}$ and $[001]_{\mathrm{LaNiGa_2}}\parallel[010]_{\mathrm{MgO}}$, or the equivalent arrangement with the LaNiGa$_2$ $[100]$ and $[001]$ axes interchanged.
These crystallographic relationships are illustrated schematically in Fig.~2.

The film thickness uniformity and surface morphology were characterized by X-ray reflectivity (XRR) and atomic force microscopy (AFM).
As shown in Fig.~1(d), the XRR profile exhibits well-defined Kiessig fringes extending over a broad angular range, indicating a uniform film thickness and relatively smooth interfaces.
The fitted reflectivity profile yields a film thickness of 53.5 nm and a roughness parameter of approximately 2.55 nm.
The AFM image in Fig.~1(e) reveals a smooth and continuous film surface with a root-mean-square roughness of 1.24 nm over a $5\times5~\mu$m$^2$ area.

The superconducting properties of the films were characterized by electrical transport measurements.
Figure~1(f) shows the temperature dependence of the resistivity, revealing a superconducting transition at $T_c\approx0.96$ K and a zero-resistance state at lower temperatures.
Although the transition temperature is reduced compared with the approximately 2 K reported for bulk LaNiGa$_2$ \cite{weng_two-gap_2016,badger_dirac_2022}, the observation of superconductivity, together with the well-defined out-of-plane orientation and narrow azimuthal diffraction peaks, demonstrates the successful growth of highly crystalline superconducting LaNiGa$_2$ thin films.

To further investigate the crystalline structure and rotational twinning revealed by XRD, cross-sectional transmission electron microscopy (TEM) measurements were performed on the LaNiGa$_2$ films.
Figure~2 shows an atomic-resolution TEM image of the LaNiGa$_2$/MgO interface, together with structural models of LaNiGa$_2$ viewed along the [100] and [001] crystallographic directions.
The TEM image reveals a highly ordered crystalline structure extending throughout the film, with well-resolved atomic columns and a distinct interface with the MgO substrate.
However, the film does not exhibit a uniform crystallographic orientation.
Instead, regions corresponding to the [100] and [001] projections coexist within the film, as identified by comparison with the simulated atomic arrangements.
These two orientations are related by a 90$^\circ$ rotation about the crystallographic $b$ axis, consistent with the rotational twinning inferred from the XRD $\phi$ scans.

Interestingly, the two crystallographic orientations are not confined to separate columnar domains.
As shown in Fig.~2, regions with different orientations can occur at different depths within the film, with boundaries extending approximately parallel to the substrate surface.
Despite these local orientation changes, the atomic ordering remains well defined, indicating that the rotational variants can be accommodated within the highly crystalline LaNiGa$_2$ film.
The coexistence of these orientations is consistent with the nearly identical lattice parameters along the $a$ and $c$ axes, which favor the formation of rotational variants during growth on the cubic MgO substrate.

Beyond the rotational twinning, atomic-resolution TEM reveals local variations in the stacking arrangements of LaNiGa$_2$.
Although the \textit{Cmcm} and \textit{Cmmm} structures have nearly identical lattice parameters, they differ in the relative lateral displacement of neighboring atomic layers, resulting in distinct crystallographic symmetries.
Figure~3(a) illustrates the atomic arrangements of the two structures viewed along the [100] and [001] directions.
The colored arrows identify characteristic atomic arrangements in successive layers, highlighting how their stacking sequences depend on both the crystal structure and crystallographic orientation.
In particular, the different projections produce distinguishable arrangements of the Ni--Ga coordination polyhedra, providing characteristic features for identifying the local structures in atomic-resolution TEM images.

Figure~3(b) presents a representative cross-sectional TEM image showing structural variations within a continuous crystalline region.
The lower portion exhibits the \textit{Cmcm} structure viewed along [100], whereas the upper portion is consistent with the \textit{Cmmm} structure viewed along [001].
Between these regions, successive layers display different atomic arrangements, as indicated by the colored arrows corresponding to the structural models in Fig.~3(a).
These variations occur on the scale of individual atomic layers, illustrating how different stacking arrangements can be accommodated within the crystalline LaNiGa$_2$ lattice.

The structural identification is further illustrated in Fig.~3(c), which shows an enlarged view of the region marked in Fig.~3(b), together with the corresponding atomic models.
A particularly distinctive feature is the arrangement of the Ni--Ga coordination polyhedra.
In the \textit{Cmmm} [001] projection, the atomic columns form an octahedral motif, whereas the \textit{Cmcm} [100] projection exhibits a split-octahedral arrangement.
These characteristic features are reproduced by the structural models overlaid on the experimental images, allowing the local atomic arrangements to be distinguished directly.
The close correspondence between the observed and modeled structures supports the identification of the different stacking configurations.

Examination of the TEM images reveals that \textit{Cmcm} is the predominant structure of the film, while \textit{Cmmm}-like arrangements occur in minority regions.
Their coexistence demonstrates that LaNiGa$_2$ can accommodate different interlayer registries within an otherwise highly ordered crystalline lattice.
This local stacking polymorphism is particularly relevant because the nonsymmorphic symmetry of \textit{Cmcm}, which is absent in \textit{Cmmm}, is responsible for the symmetry-protected electronic band degeneracies proposed for LaNiGa$_2$ \cite{badger_dirac_2022,quan_nonsymmorphic_2022}.

The observation of different stacking arrangements within a highly crystalline LaNiGa$_2$ film provides a new perspective on its crystallographic structure.
The reassignment of bulk LaNiGa$_2$ from \textit{Cmmm} to \textit{Cmcm} was established through single-crystal X-ray diffraction, where weak symmetry-sensitive reflections distinguish the two structures \cite{badger_dirac_2022}.
Our atomic-resolution TEM results reveal that the local structure can be more complex than the long-range crystallographic description.
Although \textit{Cmcm} is the dominant structure, minority \textit{Cmmm}-like arrangements occur within the same crystalline film, with changes in stacking registry taking place over only a few atomic layers.
As the two structures differ primarily through relative lateral displacements of neighboring layers, such variations can be accommodated without substantially disrupting the underlying crystalline framework.
The coexistence of different stacking arrangements therefore highlights an important distinction between the average crystal structure determined by diffraction and the local atomic arrangements.

This structural observation is particularly relevant to LaNiGa$_2$ because its proposed topological electronic properties depend critically on crystalline symmetry.
The nonsymmorphic glide and screw symmetries of \textit{Cmcm} enforce band degeneracies that produce Dirac lines and a Dirac loop at the Fermi level \cite{badger_dirac_2022,quan_nonsymmorphic_2022}.
These symmetry protections are absent in the symmorphic \textit{Cmmm} structure.
The presence of local \textit{Cmmm}-like arrangements within a predominantly \textit{Cmcm} lattice therefore raises the possibility that the electronic structure of real LaNiGa$_2$ samples is influenced by local departures from the ideal nonsymmorphic symmetry.
In particular, variations in interlayer registry may modify the electronic states associated with symmetry-enforced band degeneracies, depending on the spatial extent and distribution of the stacking variants.
Given the proposed connection between these band degeneracies and unconventional interband superconducting pairing, understanding the local stacking structure may be important for interpreting the superconducting properties of LaNiGa$_2$.

Despite the coexistence of the two structural phases, the film exhibits only a single superconducting transition at $T_C\approx0.96$~K, with a transition width ($\Delta T_C$) of approximately 0.25~K, as shown in Fig.~1(f). 
By comparison, transition widths of bulk LaNiGa$_2$ can vary from  approximately 0.025~K to 0.15~K for single crystal \cite{ghimire_electron_2024}.
The film $T_C$ is also lower than the values of approximately 2~K reported for bulk LaNiGa$_2$ assigned to either polycrystalline \textit{Cmmm} or single-crystalline \textit{Cmcm}.
A single resistive transition need not imply identical values of $T_C$ in the two structural regions, since the formation of a continuous superconducting path could mask a subsequent transition in regions with lower $T_C$.

The coexistence of \textit{Cmcm} and \textit{Cmmm} stacking arrangements may reflect an intrinsic structural tendency of LaNiGa$_2$.
Unlike the rotational twinning associated with the substrate symmetry, the distinction between these structures arises from the relative lateral displacement of neighboring atomic layers, without significant changes in lattice parameters or local atomic coordination.
First-principles calculations have shown that the two structures are energetically close, with the \textit{Cmmm} structure calculated to be 12.7 meV/atom lower in energy than \textit{Cmcm}, despite the latter being the experimentally established bulk structure \cite{quan_nonsymmorphic_2022}.
The small energy difference and close structural relationship suggest that different stacking sequences may be stabilized during crystal growth, potentially giving rise to local structural intergrowths.
Our direct observation of such intergrowths in thin films raises the possibility that similar stacking variations also occur in bulk LaNiGa$_2$.

The possibility of local structural heterogeneity is particularly relevant to the ongoing debate over the superconducting pairing symmetry of LaNiGa$_2$.
Measurements of the magnetic penetration depth and superfluid density have supported a two-band superconducting state consistent with the proposed nonunitary triplet pairing \cite{sundar_gap_2024}.
In contrast, electron irradiation experiments revealed strong robustness against nonmagnetic disorder, favoring a conventional sign-preserving pairing state \cite{ghimire_electron_2024}.
More recently, nuclear quadrupole resonance measurements reported a pronounced Hebel--Slichter coherence peak, which is consistent with two-band spin-singlet pairing but difficult to reconcile with the proposed nonunitary triplet state \cite{Sherpa2026}.
These contrasting interpretations highlight the importance of understanding the microscopic characteristics of the samples under investigation.
In particular, establishing whether local stacking polymorphism is also present in bulk LaNiGa$_2$ may provide valuable insight into the relationship between crystalline symmetry, electronic structure, and superconducting pairing.

In summary, we have demonstrated the growth of highly crystalline superconducting LaNiGa$_2$ thin films on MgO substrates, exhibiting well-defined crystallographic orientation, rotational twinning, and a superconducting transition near 1 K.
Atomic-resolution TEM reveals local stacking polymorphism, with minority \textit{Cmmm}-like arrangements embedded within the predominantly \textit{Cmcm} structure.
The coexistence of these arrangements demonstrates that distinct local crystalline symmetries can be accommodated within the same LaNiGa$_2$ lattice.
Given the essential role of nonsymmorphic symmetry in protecting the topological electronic states of LaNiGa$_2$, our findings provide new insight into its crystallographic structure and establish a superconducting thin-film platform for further investigations.

\end{text}
\bibliography{LaNiGa2_2}

@article{li_pressure-induced_2026,
	title = {Pressure-induced evolution of superconductivity and topological electronic structure in {LaNiGa} 2},
	volume = {113},
	issn = {2469-9950, 2469-9969},
	url = {https://link.aps.org/doi/10.1103/cbbg-r8ns},
	doi = {10.1103/cbbg-r8ns},
	language = {en},
	number = {10},
	urldate = {2026-06-09},
	journal = {Physical Review B},
	author = {Li, Yikang and Yang, Ye and Li, Houpu and Zhang, Yuqing and Li, Yanjun and Liu, Qingyuan and Wu, Rongqi and Cao, Xiaopei and Wang, Rui and Ying, Jianjun and Chen, Xianhui},
	month = mar,
	year = {2026},
	pages = {104505},
}

@article{ghimire_electron_2024,
	title = {Electron irradiation reveals robust fully gapped superconductivity in {LaNiGa} 2},
	volume = {109},
	issn = {2469-9950, 2469-9969},
	url = {https://link.aps.org/doi/10.1103/PhysRevB.109.024515},
	doi = {10.1103/PhysRevB.109.024515},
	language = {en},
	number = {2},
	urldate = {2026-06-09},
	journal = {Physical Review B},
	author = {Ghimire, S. and Joshi, K. R. and Krenkel, E. H. and Tanatar, M. A. and Shi, Yunshu and Kończykowski, M. and Grasset, R. and Taufour, V. and Orth, P. P. and Scheurer, M. S. and Prozorov, R.},
	month = jan,
	year = {2024},
	pages = {024515},
}

@article{sundar_gap_2024,
	title = {Gap structure of the nonsymmorphic superconductor {LaNiGa} 2 probed by μ {SR}},
	volume = {109},
	issn = {2469-9950, 2469-9969},
	url = {https://link.aps.org/doi/10.1103/PhysRevB.109.104517},
	doi = {10.1103/PhysRevB.109.104517},
	language = {en},
	number = {10},
	urldate = {2026-06-09},
	journal = {Physical Review B},
	author = {Sundar, Shyam and Yakovlev, M. and Azari, N. and Abedi, M. and Broun, D. M. and Özdemir, H. U. and Dunsiger, S. R. and Zackaria, D. and Bowman, H. and Klavins, P. and Shi, Y. and Taufour, V. and Sonier, J. E.},
	month = mar,
	year = {2024},
	pages = {104517},
}

@article{hillier_nonunitary_2012,
	title = {Nonunitary {Triplet} {Pairing} in the {Centrosymmetric} {Superconductor} {LaNiGa} 2},
	volume = {109},
	copyright = {http://link.aps.org/licenses/aps-default-license},
	issn = {0031-9007, 1079-7114},
	url = {https://link.aps.org/doi/10.1103/PhysRevLett.109.097001},
	doi = {10.1103/PhysRevLett.109.097001},
	language = {en},
	number = {9},
	urldate = {2026-07-11},
	journal = {Physical Review Letters},
	author = {Hillier, A. D. and Quintanilla, J. and Mazidian, B. and Annett, J. F. and Cywinski, R.},
	month = aug,
	year = {2012},
	pages = {097001},
}

@article{badger_dirac_2022,
	title = {Dirac lines and loop at the {Fermi} level in the time-reversal symmetry breaking superconductor {LaNiGa2}},
	volume = {5},
	copyright = {2022 The Author(s)},
	issn = {2399-3650},
	url = {https://www.nature.com/articles/s42005-021-00771-5},
	doi = {10.1038/s42005-021-00771-5},
	language = {en},
	number = {1},
	urldate = {2026-07-11},
	journal = {Communications Physics},
	publisher = {Nature Publishing Group},
	author = {Badger, Jackson R. and Quan, Yundi and Staab, Matthew C. and Sumita, Shuntaro and Rossi, Antonio and Devlin, Kasey P. and Neubauer, Kelly and Shulman, Daniel S. and Fettinger, James C. and Klavins, Peter and Kauzlarich, Susan M. and Aoki, Dai and Vishik, Inna M. and Pickett, Warren E. and Taufour, Valentin},
	month = jan,
	year = {2022},
	pages = {22},
}

@article{zeng_superconductivity_2002,
	title = {Superconductivity in the {Ni}-based ternary compound {LaNiGa} 2},
	volume = {66},
	copyright = {http://link.aps.org/licenses/aps-default-license},
	issn = {0163-1829, 1095-3795},
	url = {https://link.aps.org/doi/10.1103/PhysRevB.66.092503},
	doi = {10.1103/PhysRevB.66.092503},
	language = {en},
	number = {9},
	urldate = {2026-07-11},
	journal = {Physical Review B},
	author = {Zeng, N. L. and Lee, W. H.},
	month = sep,
	year = {2002},
	pages = {092503},
}

@article{weng_two-gap_2016,
	title = {Two-{Gap} {Superconductivity} in {LaNiGa} 2 with {Nonunitary} {Triplet} {Pairing} and {Even} {Parity} {Gap} {Symmetry}},
	volume = {117},
	copyright = {http://link.aps.org/licenses/aps-default-license},
	issn = {0031-9007, 1079-7114},
	url = {https://link.aps.org/doi/10.1103/PhysRevLett.117.027001},
	doi = {10.1103/PhysRevLett.117.027001},
	language = {en},
	number = {2},
	urldate = {2026-07-11},
	journal = {Physical Review Letters},
	author = {Weng, Z. F. and Zhang, J. L. and Smidman, M. and Shang, T. and Quintanilla, J. and Annett, J. F. and Nicklas, M. and Pang, G. M. and Jiao, L. and Jiang, W. B. and Chen, Y. and Steglich, F. and Yuan, H. Q.},
	month = jul,
	year = {2016},
	pages = {027001},
}

@article{ghosh_quantitative_2020,
	title = {Quantitative theory of triplet pairing in the unconventional superconductor {LaNiGa} 2},
	volume = {101},
	issn = {2469-9950, 2469-9969},
	url = {https://link.aps.org/doi/10.1103/PhysRevB.101.100506},
	doi = {10.1103/PhysRevB.101.100506},
	language = {en},
	number = {10},
	urldate = {2026-07-11},
	journal = {Physical Review B},
	author = {Ghosh, Sudeep Kumar and Csire, Gábor and Whittlesea, Philip and Annett, James F. and Gradhand, Martin and Újfalussy, Balázs and Quintanilla, Jorge},
	month = mar,
	year = {2020},
	pages = {100506},
}

@article{quan_nonsymmorphic_2022,
	title = {Nonsymmorphic band sticking in a topological superconductor},
	volume = {105},
	issn = {2469-9950, 2469-9969},
	url = {https://link.aps.org/doi/10.1103/PhysRevB.105.064517},
	doi = {10.1103/PhysRevB.105.064517},
	language = {en},
	number = {6},
	urldate = {2026-07-11},
	journal = {Physical Review B},
	author = {Quan, Yundi and Taufour, Valentin and Pickett, Warren E.},
	month = feb,
	year = {2022},
	pages = {064517},
}

@article{li_fermiology_2025,
	title = {Fermiology with nodal structures in nonsymmorphic superconductor {LaNiGa2}: {A} de {Haas}-van {Alphen} study},
	volume = {68},
	issn = {1674-7348, 1869-1927},
	shorttitle = {Fermiology with nodal structures in nonsymmorphic superconductor {LaNiGa2}},
	url = {https://link.springer.com/10.1007/s11433-024-2559-x},
	doi = {10.1007/s11433-024-2559-x},
	language = {en},
	number = {4},
	urldate = {2026-07-12},
	journal = {Science China Physics, Mechanics \& Astronomy},
	author = {Li, Houpu and Yang, Ye and Shi, Mengzhu and Qian, Yingcai and Pan, Senyang and Fan, Kaibao and Zhang, Nan and Tang, Kaixin and Li, Hongyu and Wang, Zhiwei and Zhang, Jinglei and Xi, Chuanying and Xiang, Ziji and Chen, Xianhui},
	month = apr,
	year = {2025},
	pages = {247414},
}

@article{staab_symmetry-enforced_2024,
	title = {Symmetry-enforced {Fermi} surface degeneracies observed in the purported time-reversal symmetry-broken superconductor {LaNiGa} 2},
	volume = {110},
	issn = {2469-9950, 2469-9969},
	url = {https://link.aps.org/doi/10.1103/PhysRevB.110.165115},
	doi = {10.1103/PhysRevB.110.165115},
	language = {en},
	number = {16},
	urldate = {2026-07-13},
	journal = {Physical Review B},
	author = {Staab, Matthew and Prater, Robert and Sreedhar, Sudheer and Byland, Journey and Mann, Eliana and Zackaria, Davis and Shi, Yunshu and Bowman, Henry J. and Stephens, Andrew L. and Jung, Myung-Chul and Botana, Antia S. and Pickett, Warren E. and Taufour, Valentin and Vishik, Inna},
	month = oct,
	year = {2024},
	pages = {165115},
}

@Article{Sherpa2026,
  author    = {Sherpa, P. and Hingorani, R. and Menon, A. and Vinograd, I. and Chaffey, C. and Dioguardi, A. P. and Yamamoto, R. and Hirata, M. and Ronning, F. and Badger, J. R. and Klavins, P. and Singh, R. R. P. and Taufour, V. and Curro, N. J.},
  journal   = {Physical Review Letters},
  title     = {Observation of a Pronounced Hebel--Slichter Peak in the Spin-Lattice Relaxation Rate and Implications for Gap and Pairing Symmetry in {LaNiGa$_2$}},
  year      = {2026},
  issn      = {1079-7114},
  month     = May,
  number    = {18},
  volume    = {136},
  pages     = {186001},
  doi       = {10.1103/tsnb-1mxc},
  publisher = {American Physical Society (APS)},
}

\clearpage
\begin{figure}
\centering
\includegraphics[width=16cm]{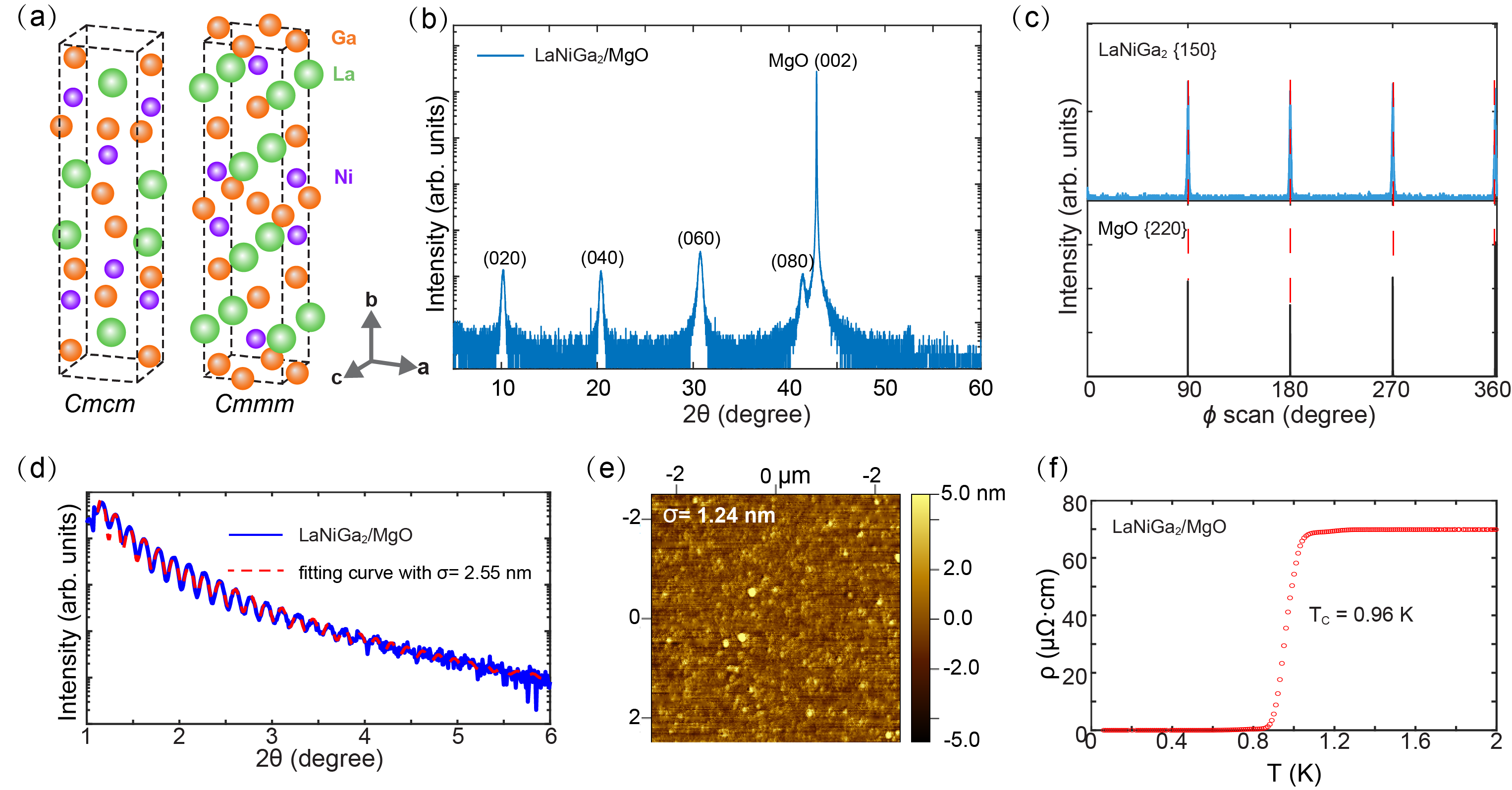}
\caption{ \textbf{Structural characterization and superconductivity of LaNiGa$_2$ thin films grown on MgO(001).} (a) Schematic crystal structures of LaNiGa$_2$ in the \textit{Cmcm} and \textit{Cmmm} space groups. La, Ni and Ga atoms are shown in green, purple and orange, respectively. (b) Out-of-plane X-ray diffraction pattern of the LaNiGa$_2$/MgO sample. (c) In-plane $\phi$ scans of the LaNiGa$_2$\{150\} and MgO\{220\} reflections. The aligned peak positions indicate a defined in-plane orientation relationship between the film and substrate. (d) X-ray reflectivity data (blue) and fit (red dashed line), yielding a fitted roughness of $\sigma$ =2.55 nm. (e) Atomic force microscopy image of the film surface, with a root-mean-square roughness of 1.24 nm. (f) Resistivity measurement showing a superconducting transition at \textit{Tc}=0.96 K.}
\end{figure}

\begin{figure}
\centering
\includegraphics[width=16cm]{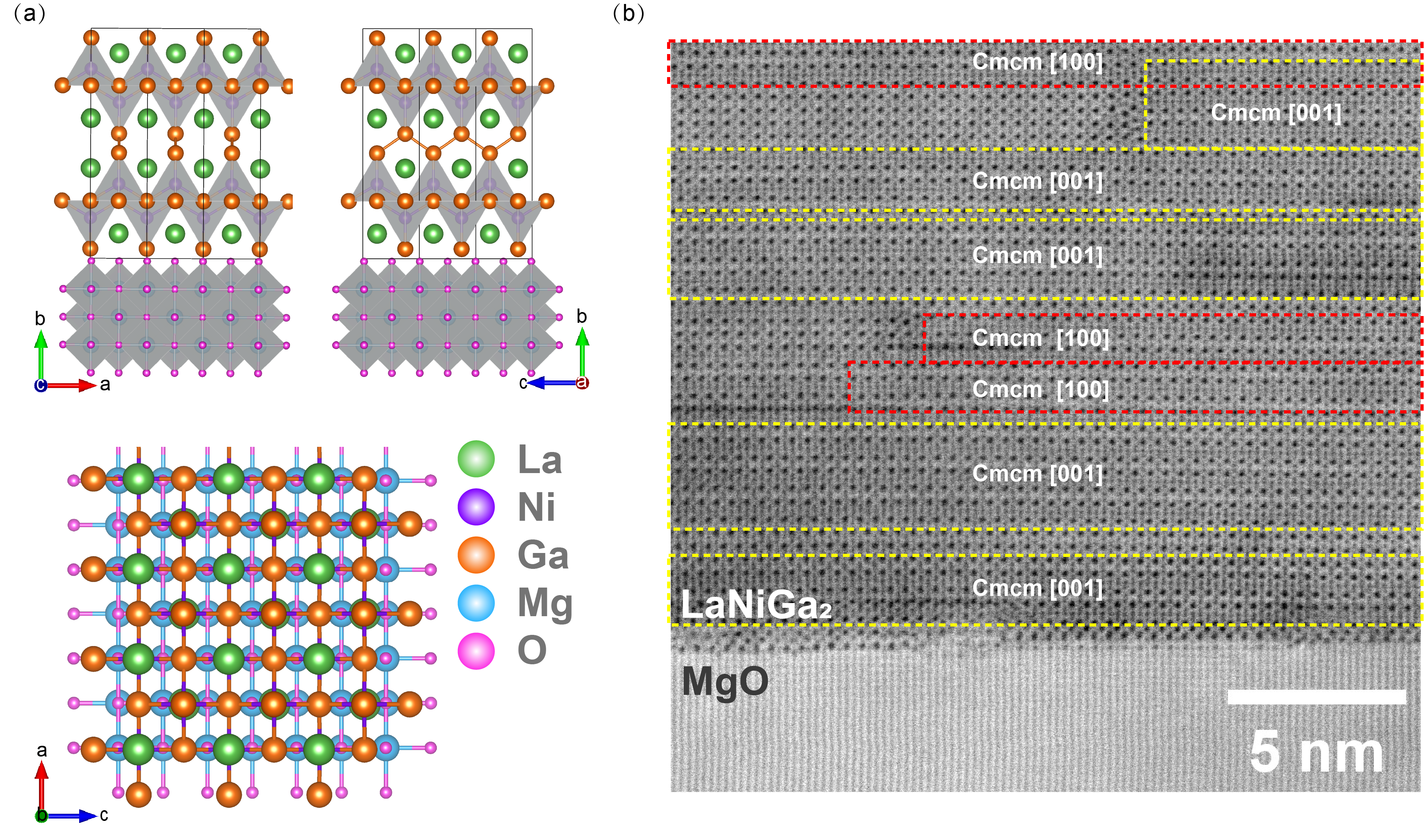}
\caption{ \textbf{Crystal structure and epitaxial growth of LaNiGa$_2$ thin films on MgO.} (a) Schematic crystal structures of orthorhombic LaNiGa$_2$ (space group \textit{Cmcm}) on the MgO substrate, viewed along the [001](top left) and [100](top right) zone axes, with the b axis of LaNiGa$_2$ along the out-of-plane direction. The bottom panel shows the plane view along the b axis, illustrating the in-plane lattice matching between LaNiGa$_2$ and MgO. (b) Cross-sectional high-resolution STEM image of a LaNiGa$_2$ film on MgO, showing a twinned domain structure. The film consists of two types of twin domains related by a 90° rotation about the out-of-plane b axis, which are viewed along the [001](yellow dashed boxes) and [100] (red dashed boxes) of LaNiGa$_2$, respectively. The twin domains are stacked alternately along the growth direction. Scale bar: 5 nm.}
\end{figure}

\begin{figure}
\centering
\includegraphics[width=16cm]{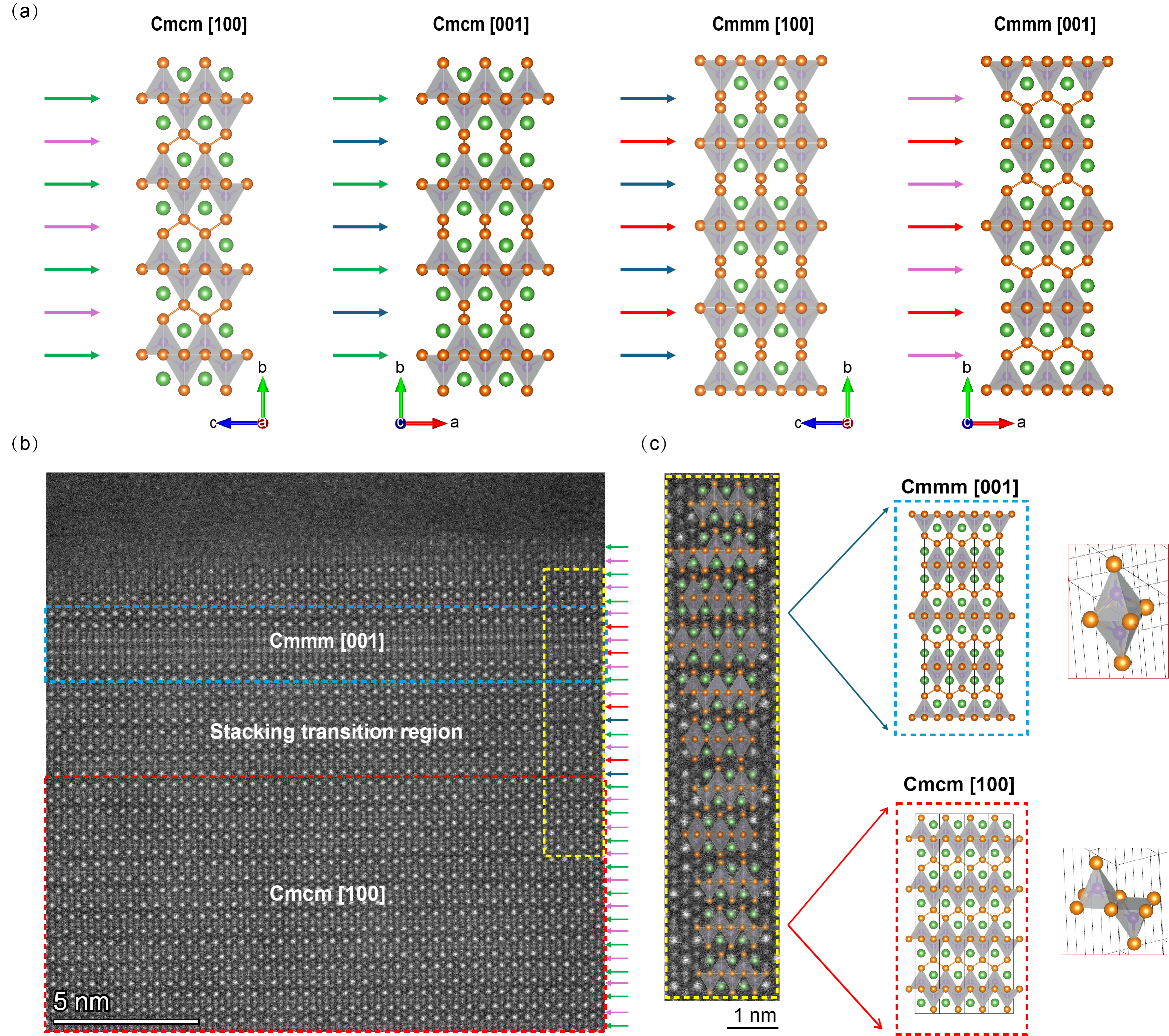}
\caption{ \textbf{Local stacking polymorphism and atomic arrangements in LaNiGa$_2$ thin films.} (a) Crystallographic models of \textit{Cmcm} and \textit{Cmmm} viewed along the [100] and [001] directions. Colored arrows identify characteristic atomic arrangements in successive layers. (b) Cross-sectional atomic-resolution STEM image showing local structural variations. Blue and red dashed boxes identify regions corresponding to \textit{Cmmm} [001] and \textit{Cmcm} [100], respectively. The intervening region exhibits variations in the stacking sequence, indicated by colored arrows corresponding to the structural models in (a). Scale bar: 5 nm. (c) Enlarged view of the yellow-boxed region in (b), with atomic models overlaid on the experimental image. The corresponding structural models and coordination polyhedra highlight the octahedral motif in the \textit{Cmmm} [001] projection and the split-octahedral arrangement in the \textit{Cmcm} [100] projection.
}
\end{figure}

\clearpage

\textbf{ACKNOWLEDGMENTS}

This work was supported by National Science Foundation of China (Grant No. 12474153), Guangdong Provincial Quantum Science Strategic Initiative (No. GDZX2301007, No. GDZX2401003), Research Grants Council, University Grants Committee of the Hong Kong SAR (Grant No. CUHK 24302822, No. CUHK 14306823), CUHK Direct Grants (No. 4053668, No. 4053727).
\end{document}